\documentclass[aps,prl, reprint]{revtex4-1}
\bibpunct{}{}{,}{s}{}{} 
\usepackage{graphicx}
\usepackage{dcolumn}
\usepackage{bm}
\usepackage{amsmath}
\usepackage{amssymb}
\usepackage{xcolor}
\usepackage[normalem]{ulem}
\usepackage{epstopdf}

\begin{document}
\title{Absorption of High Intensity, High Contrast Femtosecond Laser Pulses by a Solid}
\author{Amitava Adak}\email{adak.amitava@gmail.com[present affiliation: JILA, University of Colorado at Boulder, Boulder, Colorado 80309-0440, USA]}
\author{Amit D. Lad, Moniruzzaman Shaikh, Indranuj Dey, Deep Sarkar}
 \author{G. Ravindra Kumar}  \email{grk@tifr.res.in}
 \affiliation{Tata Institute of Fundamental Research, Dr. Homi Bhabha Road, Colaba, Mumbai-400005, India}

\begin{abstract}
The basic understanding of high-intensity femtosecond laser absorption in a solid is crucial for high-energy-density science. This multidimensional problem has many variables like laser parameters, solid target material, and geometry of the excitation. This is important for a basic understanding of intense laser-matter interaction as well for applications such as `plasma mirror'.  Here, we have experimentally observed high-intensity, high-contrast femtosecond laser absorption by an optically polished fused silica target at near-relativistic laser intensities ($\sim$10$^{18}$ W/cm$^2$). The laser absorption as a function of angle of incidence and incident energy is investigated for both $p$- and $s$-polarized pulses in detail, providing a strong indication of the presence of collisionless processes. At an optimum angle of incidence, almost as large as 80\% of the laser ($p$-polarized) energy gets absorbed in the target. Such a high percentage of absorption at near-relativistic intensities has not been observed before. At smaller angles of incidence the high reflectivity (e.g. about 60 - 70\% at 30$^\circ$ incidence) indicate that, this study is fundamentally relevant for plasma mirrors at near-relativistic intensities.
\end{abstract}
\maketitle

A thorough investigation of a high-intensity laser absorption in a solid target is important for understanding the basic processes and their consequences\cite{Gibbonbook,WilksPRL1992,WilksIEEE1997,PingPRL2008,PrashantSciRep2015}. Various mechanisms of high intensity short pulse laser absorption (resonance absorption, vacuum heating, anomalous skin effect, $\vec{J}\times \vec{B}$ heating etc.) by solids have been predicted, explored with numerical simulations, analytical works and experimental studies\cite{Kruerbook,BrunnelPRL1987,AndreevJETP1992,GibbonPRL1992,FedosejevsPRL1990}. All these absorption mechanisms related to high intensity short pulse lasers are very sensitive to several experimental parameters, such as (i) laser intensity at target surface -- the local intensity of the laser on the target surface can be optimized by various target designs\cite{PrashantSINW2012,RajeevPRL2003}, (ii) preplasma scale length -- which depends on prepulse intensity contrast of the laser, (iii) angle of incidence and polarization of the laser. In a typical experiment, extracting the contribution of a particular mechanism is often difficult.

Vacuum heating is one of the key processes of intense laser absorption, where an obliquely incident intense electromagnetic wave on a sharply bounded overdense plasma drags the electrons into the vacuum, and sends back into the overdense plasma (first identified by Brunel\cite{BrunnelPRL1987}). Thereafter, several of numerical simulations\cite{GibbonPRL1992,Bulanov1994,RuhlPLA1995} were used to predict this process until the first experimental observation was reported by Grimes \textit{et al.}\cite{GrimesVH} at non-relativistic intensities. This process needs a preplasma scale length smaller than the quiver amplitude of the electrons in the presence of the laser field. However, this process at near-relativistic intensities still remains unexplored.

A previous study by Ping \textit{et al}\cite{PingPRL2008} using 150 fs laser pulses showed more than 80\% absorption beyond the intensities of 10$^{20}$ W/cm$^{2}$ by combined effect of `large preplasma' and `hole-boring'. At oblique incidence the dependence of the laser absorption at near-relativistic intensities (around 10$^{18}$ W/cm$^2$) was not clear due to noise in the signal. Also the experiment was performed only at two angle of incidence. Another study by J. R. Davies\cite{Davies2009} shows the femtosecond laser absorption and simulation across a broad intensity range up to beyond 10$^{20}$ W/cm$^2$. However, none of the experimental observation capture a clear behaviour in the near-relativistic intensity regime. 

In this study, we investigate near-relativistic intensity ($\sim$ 10$^{18}$ W/cm$^2$), high contrast ($\sim$10$^{-9}$), 30 fs laser absorption in a fused silica target. Dependence of the absorption on angle of incidence and incident laser intensity is explored in detail. An indication of a strong vacuum heating, particularly at much oblique incidence, is observed. Absorption as high as 80\% of the
laser energy is seen for the first time at near-relativistic intensities. On the other hand, at smaller angle of incidences the high reflectivity (e.g. about 60 - 70\% at 30$^\circ$ incidence) indicates that this study is relevant in the research of plasma mirror at these intensities\cite{TsaiPOP2015}.
\section{Experimental Setup}
\begin{figure}[h]
\begin{center}
\includegraphics[scale=0.4]{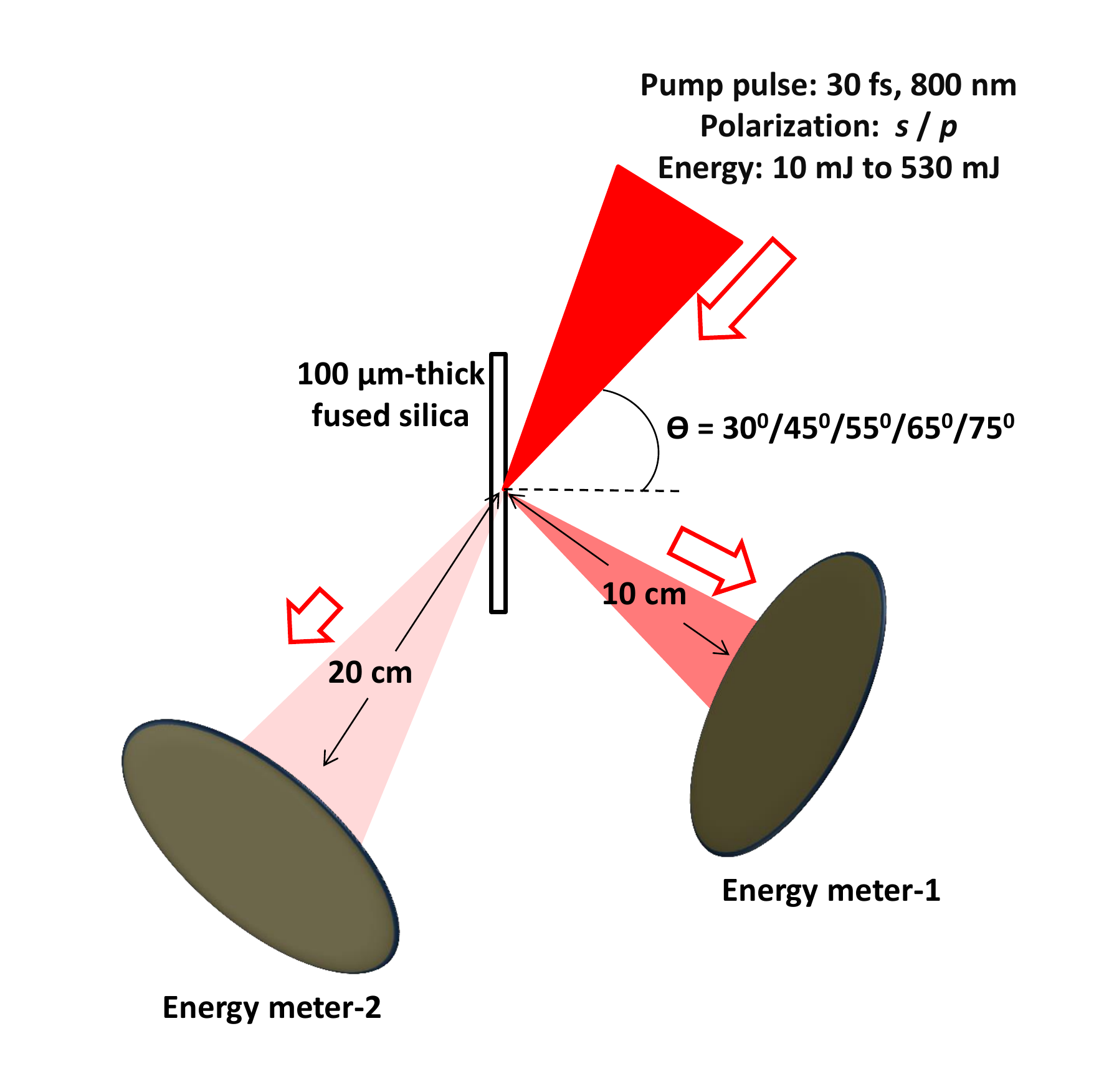}
\caption{Experimental setup: A 100 $\mu$m-thick fused silica target was irradiated by a high contrast intense femtosecond laser (800 nm, 30 fs, $I\sim6\times10^{16}$-$5\times10^{18}$ W/cm$^2$). The reflected and transmitted lasers were collected by energy meter-1 and 2 respectively. The angle of incidence ($\theta$) was varied from 30$^{\circ}$ to 75$^{\circ}$ for both the case of \textit{p}- ans \textit{s}-polarization.}\label{setup}
\end{center}
\end{figure}

The experiment was carried out using a high intensity contrast ($\sim$10$^{-9}$), high power Ti:Sapphire laser system (100 TW, 800 nm, 30 fs) at Ultrashort Pulse High Intensity Laser Laboratory (UPHILL) at Tata Institute of Fundamental Research, Mumbai. The laser was focused by a gold coated off-axis parabolic mirror to a spot of $\sim$11 $\mu$m (FWHM) on a plane 100 $\mu$m-thick fused silica target kept inside 10$^{-6}$ Torr vacuum chamber. Two energy meters were used to measure the reflected and transmitted laser energy and thereby the absorbed laser energy (non-specular scattering $<2$\% of input). The solid angles covered by the energy meter sensors (pyroelectric, OPHIR PE100BF-DIF-C) at the interaction point, were a few times larger than those for the specular laser reflection (see Fig. \ref{setup}). Transmission-calibrated glass-debris-shields were placed in front of those energy meters to protect them from energetic particles. We repeated this measurement at various angle of incidence with a range (10 to 530 mJ) of incident energies both for \textit{p}- and \textit{s}-polarization of the laser. For each experimental condition, data were taken over several laser shots to minimize the uncertainty (error) in the measured absorption.
 
\section{Results and discussion}

 \begin{figure}[h]
\begin{center}
\includegraphics[width=\columnwidth]{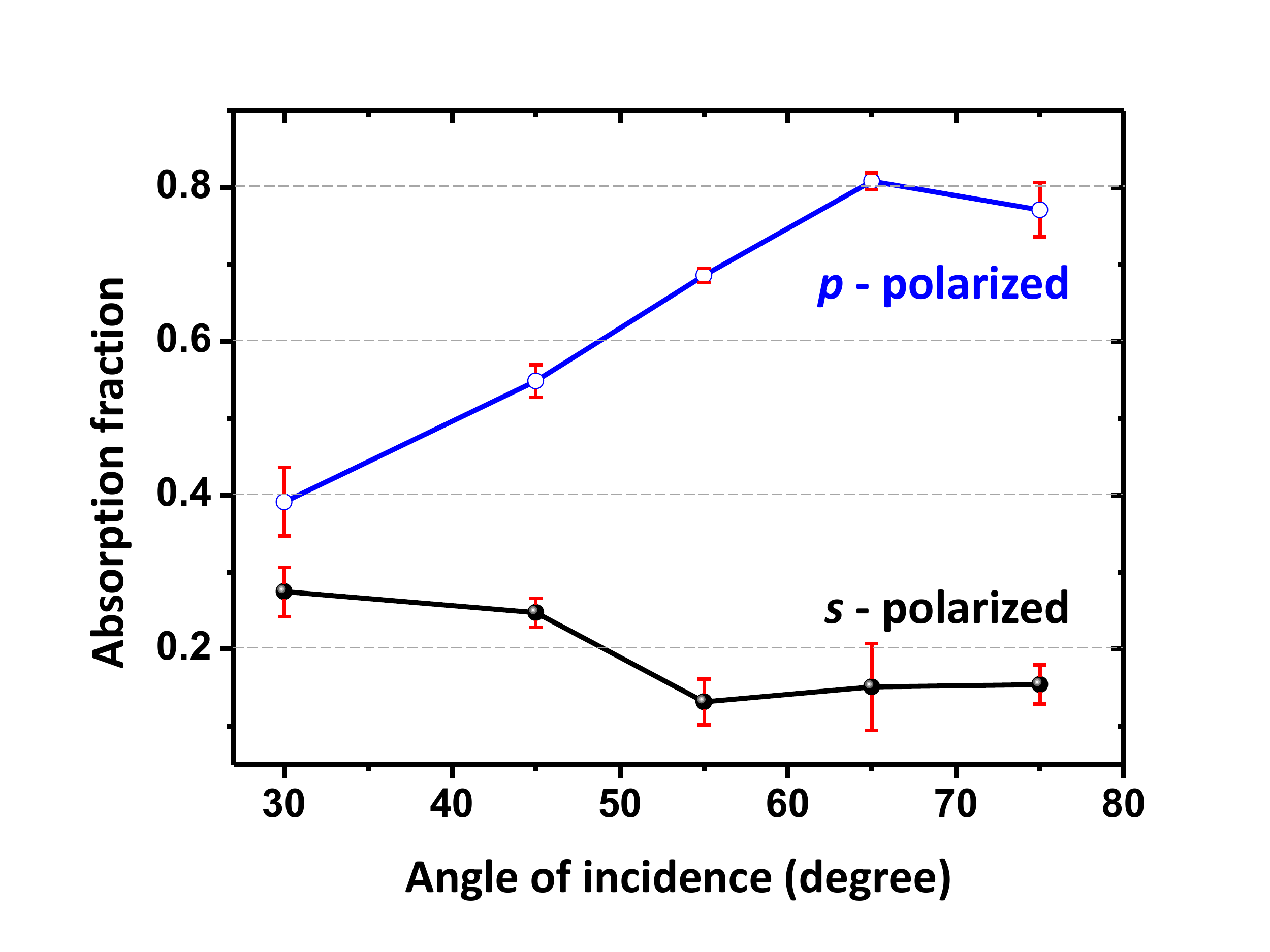} 
\caption{Absorption vs angle of incidence for \textit{p}- and \textit{s}-polarized high contrast femtosecond laser at 230 $\pm$ 10 mJ incident energy (this energy at 45$^{\circ}$ angle of incidence corresponds to the laser intensity of $I\sim1.8\times10^{18}$ W/cm$^2$).}\label{AvsAOI}
\end{center}
\end{figure}

Figure \ref{AvsAOI} shows the variation of laser absorption as a function of angle of incidence for both \textit{p}- and \textit{s}-polarized laser with (230 $\pm$ 10) mJ incident energy. The focused laser intensity at the target surface in this case was $I_L\gtrsim10^{18}$ W/cm$^2$. The absorption for \textit{p}-polarized laser increases from $\sim$ 40\% at 30$^{\circ}$ angle of incidence to $\sim$ 80\% at 65$^{\circ}$. Beyond this angle, the absorption tends to decrease. On the other hand, for \textit{s}-polarized laser the absorption decreases from $\sim$ 25\% at 30$^{\circ}$ angle of incidence to $\sim$ 15\% at 55$^{\circ}$. Beyond this angle, the absorption always remains low. This result of increasing separation between the absorption curves (for \textit{p}- and \textit{s}-polarization) with increase in angle of incidence strongly indicates dominant collisionless collective absorption processes for near-relativistic intensity pulses and is particularly prominent at higher angles of incidence.

\begin{figure}[t]
\begin{center}
\includegraphics[width=\columnwidth]{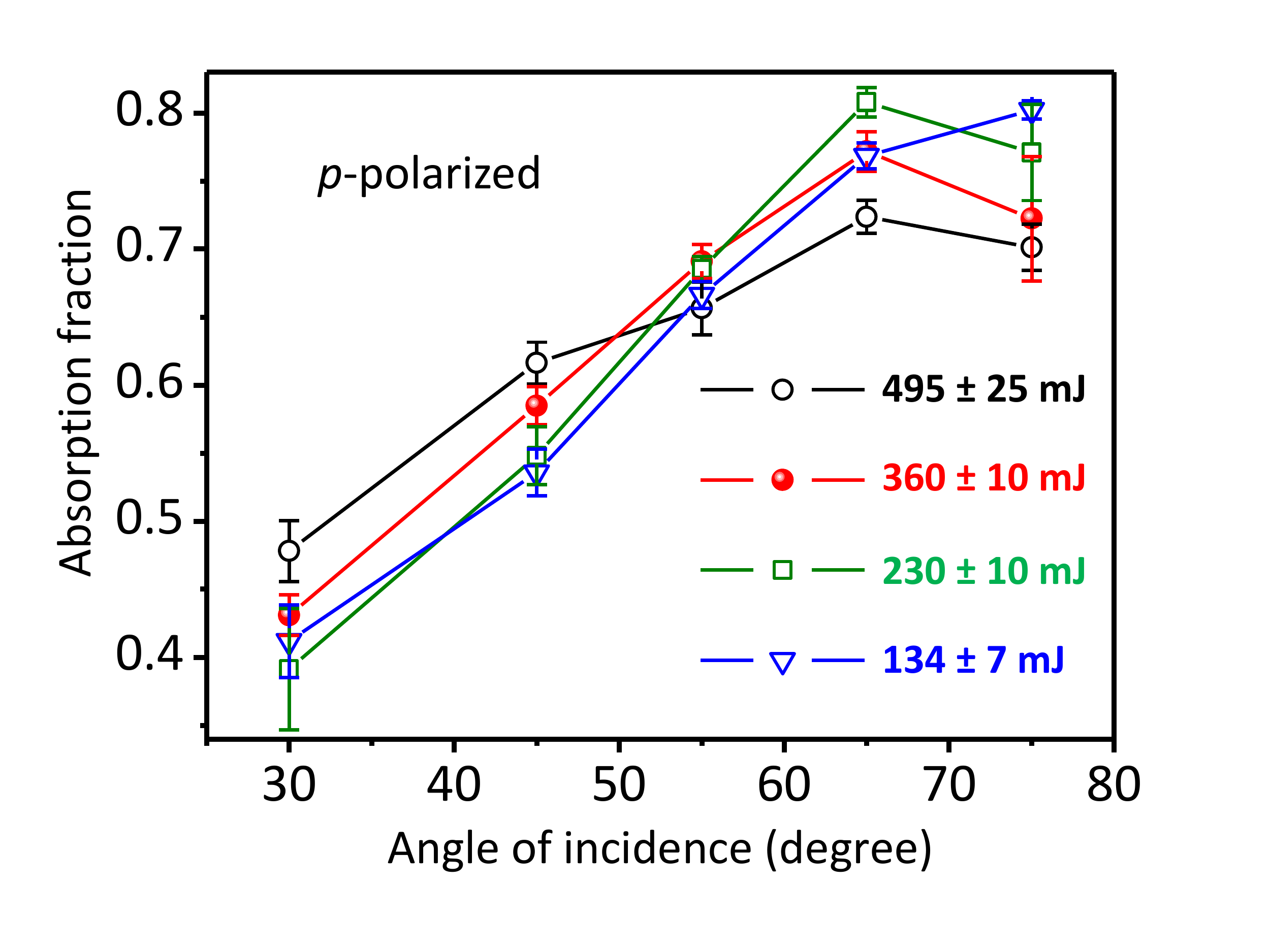}
\caption{Comparison of absorption as a function of angle of incidence for \textit{p}-polarization of the incident laser with energies of 134 $\pm$ 7, 230 $\pm$ 10, 360 $\pm$ 10, and 495 $\pm$ 25 mJ. 230 mJ at 45$^{\circ}$ angle of incidence corresponds to the laser intensity of $I\sim1.8\times10^{18}$ W/cm$^2$.}\label{ApvsAOIallE}
\end{center}
\end{figure}

The absorption of \textit{p}-polarized laser as a function of angle of incidence is plotted in Fig. \ref{ApvsAOIallE}, at various incidence energies up to 100s of mJ. From 40 - 50\% level at 30$^{\circ}$, the absorption increases to 70 - 80\% level at 65$^{\circ}$ angle of incidence. The absorption decreases from 65$^{\circ}$ to 75$^{\circ}$ angle of incidence except for the case of 134 mJ energy, where it shows less rate of increase instead of a decrease. 

 \begin{figure*}[]
\begin{center}
\includegraphics[width=\textwidth]{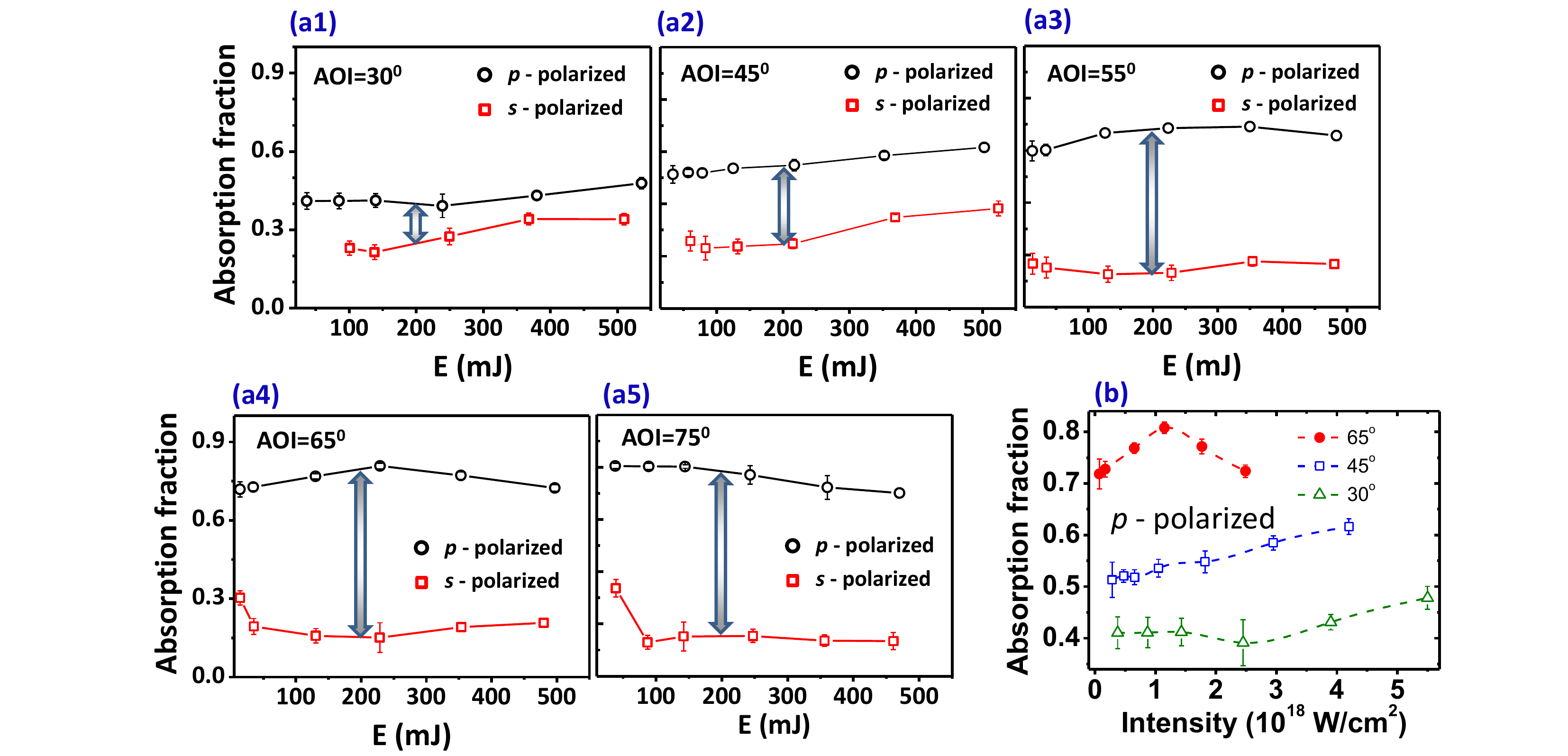}
\caption{(a1-a5) The variation of absorption for both the \textit{p}- and \textit{s}-polarization as a function of intensity at 30$^{\circ}$, 45$^{\circ}$, 55$^{\circ}$, 65$^{\circ}$, and 75$^{\circ}$ angle of incidences. The arrows indicate the separation of the absorption curves for \textit{p}- and \textit{s}-polarization. (b) Absorption of \textit{p}-polarized laser with intensity at three angle of incidences- 30$^{\circ}$, 45$^{\circ}$, 65$^{\circ}$}\label{AvsE}
\end{center}
\end{figure*}

 Figure \ref{AvsE} (a1 to a5) compares the variation of the absorption as a function of incident laser energy in both laser polarizations for various angle of incidences (30$^{\circ}$, 45$^{\circ}$, 55$^{\circ}$, 65$^{\circ}$, and 75$^{\circ}$, respectively). Almost for all laser energies, the separation between the absorption curves (for \textit{p}- and \textit{s}-polarization) increases with the angle of incidence.  The large separation between the absorption curves in \textit{p} and \textit{s}- polarized laser, particularly at higher angle of incidence, indicates strong collisionless contributions in this entire laser energy range.
Fig. \ref{AvsE}(b) shows the variation of absorption of \textit{p}-polarized pulses at near-relativistic intensities for three different angle of incidences,  30$^{\circ}$ (green triangles), 45$^{\circ}$ (blue squares), 65$^{\circ}$ (red circles). The absorption, being almost constant for 30$^{\circ}$ angle of incidence around $\sim10^{18}$ W/cm$^2$, increases slowly towards relativistic intensities. For 45$^{\circ}$ angle of incidence, the absorption keeps on increasing across this whole intensity range. On the other hand, for  65$^{\circ}$ angle of incidence, the absorption first increases (more steeply compared to former cases) upto $\sim10^{18}$ W/cm$^2$ and then decreases rapidly. We think that for 65$^{\circ}$ angle of incidence, where the vacuum heating could be the dominant absorption mechanism, increase in laser intensity changes the density scale length to go beyond electron oscillation amplitude. For other angles this reversal could be at much higher intensities which we did not observe in our experiment. However more detailed experiments and simulations are required to confirm this hypothesis.

The oscillation amplitude of the electrons driven by the field of near-relativistic intensity laser ($I=3\times10^{18}$ W/cm$^2$) can be estimated as $(\frac{X_{osc}}{\lambda})\sim 0.2$. The laser absorption by vacuum heating is efficient for $(\frac{L}{\lambda})<(\frac{X_{osc}}{\lambda})$ which is expected to be the case for our experimental regime. 
\begin{figure}[b]
\begin{center}
\includegraphics[width=\columnwidth]{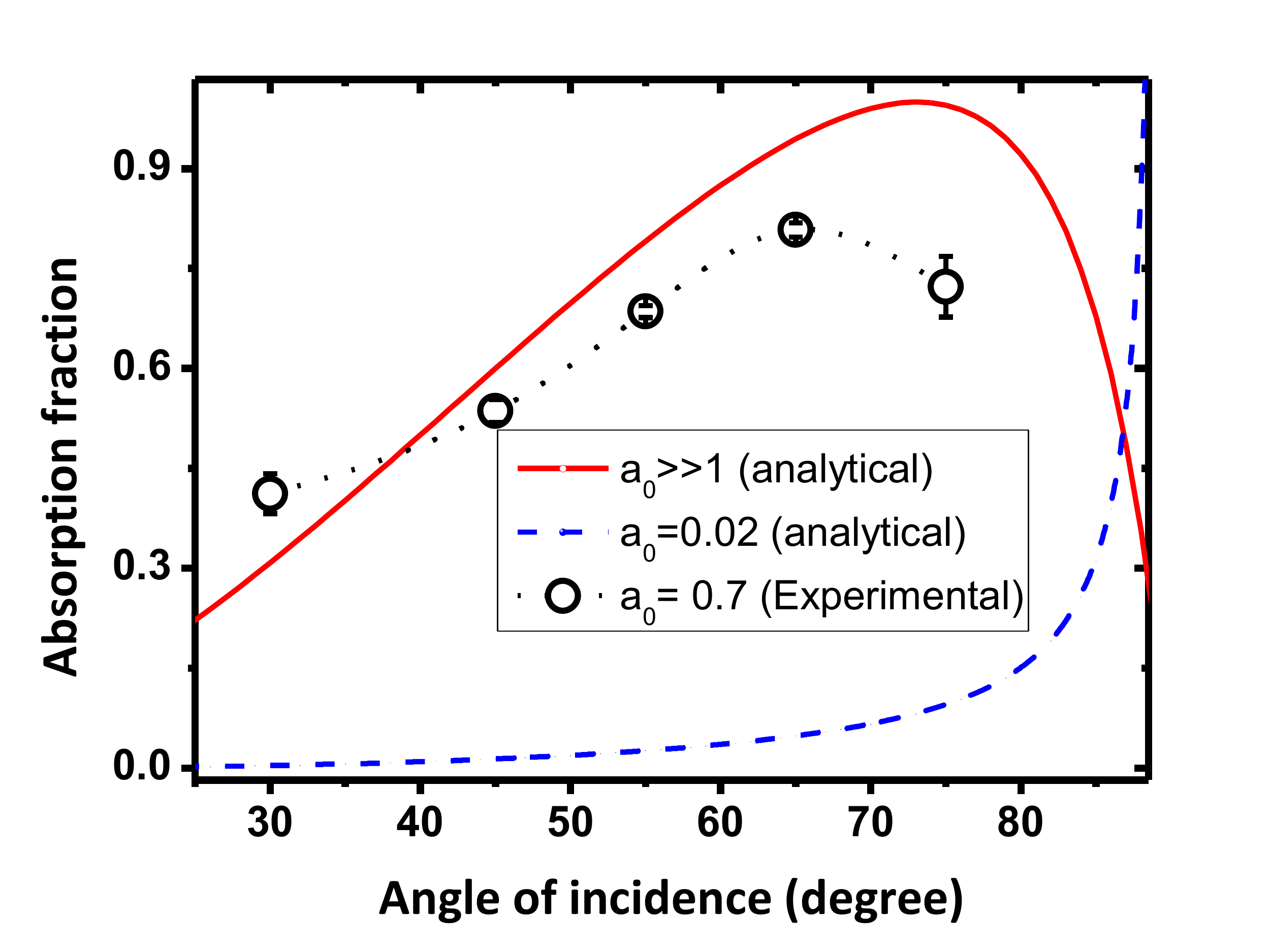}
\caption{Analytical model (Brunnel's model and its relativistic correction) of Vacuum heating vs angle of incidence for $a_{0}<<1$ (blue-dashed) and $a_{0}>>1$ (red-solid) are shown. Experimental observation of absorption fraction at $a_{0}=0.7$ is also shown (black circle).}\label{AnalytExp}
\end{center}
\end{figure}
 The black circles in Fig. \ref{AnalytExp} are the experimental results for the laser intensity corresponding to $a_0\sim0.7$, whereas blue dashed line shows the well known (Brunel) non relativistic scaling for vacuum heating as a function of angle of incidence at $a_0=0.02$. The red solid line in the same plot indicates the variation of the vacuum heating absorption efficiency for highly relativistic intensity ($a_0>>1$)\cite{Gibbonbook,BrunnelPRL1987}. The scaling with angle of incidence in the experiment indicates the presence of vacuum heating process. Additionally there could be some contribution from anomalous skin effect. However, the anomalous skin effect should monotonically increase with angle of incidence until 90$^{\circ}$ angle of incidence except in a small interval around it\cite{AndreevJETP1992}. 
 
 However, it is difficult to compare of our experimental results with these simple analytical models. For an instance, analytic Brunel heating estimates assume that the preplasma conditions are not substantially changed by the interaction of the laser pulse with the plasma, which may not be correct. In relatively recent work\cite{CaiPOP2006}, proposed an analytical fluid model for vacuum heating during the oblique incidence by an ultrashort ultraintense \textit{p}-polarized laser on a solid-density plasma. It is shown that, the optimum angle of incidence for maximum laser absorption moves towards higher values for higher incident intensities. However, much lower absorption coefficients are found. Absorption was most efficient at $a_{L}\sim$ 1.5 for an optimum angle around $45^{\circ}$-$52^{\circ}$. The electron density at the target surface is usually found to be significantly altered by the laser ponderomotive force,\cite{CaiPOP2006} an effect not considered in the earlier studies. Cai $et$ $al.$\cite{CaiPOP2006} also showed that absorption rate increases with density scale length.

\section{Relevance of present study for the plasma mirrors}
 \begin{figure}[h]
\begin{center}
\includegraphics[width=\columnwidth]{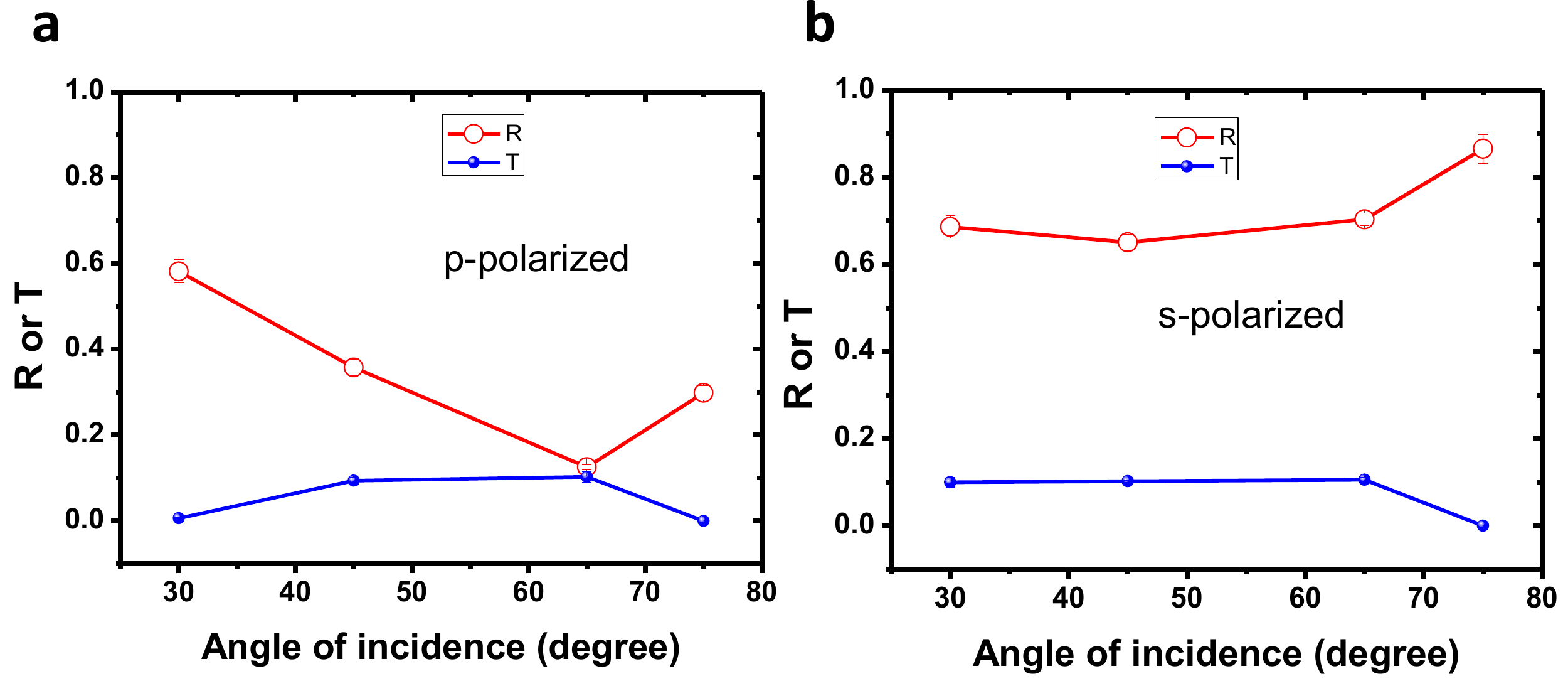}
\caption{The reflectivity (R) and the transmissivity (T) are plotted as a function of angle of incidence for (a) \textit{p}-polarization and (b) \textit{s}-polarization at the incident laser intensity of (1.6 $\pm$ 0.2)$\times10^{18}$ W/cm$^2$.}\label{Rfl_trans}
\end{center}
\end{figure}

Figure \ref{Rfl_trans} shows that, the pump reflectivity for the \textit{s}-polarization are very high (more than 65\%) for all angles of incidence. At 75$^\circ$ angle of incidence it reaches around 90\%. The pump transmission for the \textit{s}-polarization remains at or below 10\%. Whereas for the \textit{p}-polarization, the reflectivity reaches a minimum at 65$^\circ$ angle of incidence and keeps on increasing towards lower angle of incidences (e.g. about 60\% at 30$^\circ$). The pump transmission for this case remains around or below 15\%. The high reflectivity for both the polarizations at lower angles of incidence indicates that, this study is fundamentally important towards achieving an efficient plasma mirror at relativistic intensities.

\section{Summary}
In summary, absorption of a near-relativistic intensity high contrast laser by a fused silica target is explored in detail. We have observed the dependence of the absorptivity on angle of incidence and incident laser energy. A strong indication of vacuum heating process, particularly at highly oblique incidences ($55^\circ$-$75^\circ$), is experimentally observed. For the first time, absorption as high as 80\% of the laser energy is experimentally shown to be possible at near-relativistic intensities. Furthermore, the high reflectivity for both the polarization towards lower angle of incidences indicates that, this study is fundamentally important for the physics and application of plasma mirror at near-relativistic intensities. For a more clear understanding of our observations, simulations are needed.

\section{Acknowledgement}

GRK  acknowledges a  J. C. Bose Fellowship grant (JCB-037/2010).

\section{Author contributions} AA and GRK conceptualized the experiment. AA and ADL designed the experimental setup and performed the experiments with the help of all other authors. AA and ADL analysed the data and wrote the paper discussing with all other authors.

\end{document}